\documentclass[preprint,times]{elsarticle}

\usepackage[margin=2.5cm]{geometry}
\usepackage{hyperref}
\usepackage{tabularx}
\usepackage{float}
\usepackage{subfig}
\usepackage{xspace}
\usepackage{multirow}
\usepackage{makecell}
\usepackage{amsmath,amssymb}
\usepackage{booktabs}
\usepackage{lineno}
\usepackage{placeins}
\biboptions{sort&compress}

\newcommand{\abb}{\mathrm{ab^{-1}}}
\newcommand{\lumismall}{\ensuremath{100\,\mathrm{fb^{-1}}}\xspace}
\newcommand{\lumilarge}{\ensuremath{1\,\mathrm{ab^{-1}}}\xspace}
\newcommand{\mjj}{m_{jj}}
\newcommand{\logloss}{\ensuremath{\log(\text{loss})}\xspace}
\newcommand{\ycut}{\ensuremath{|y^{*}| < 0.6}\xspace}
\newcommand{\SigPreLowY}{\ensuremath{7.3\,\sigma}\xspace}
\newcommand{\SigPreLowAD}{\ensuremath{9.7\,\sigma}\xspace}
\newcommand{\SigPostLowY}{\ensuremath{3.4\,\sigma}\xspace}
\newcommand{\SigPostLowAD}{\ensuremath{5.8\,\sigma}\xspace}
\newcommand{\SigPreHighY}{\ensuremath{23.2\,\sigma}\xspace}
\newcommand{\SigPreHighAD}{\ensuremath{30.8\,\sigma}\xspace}
\newcommand{\SigPostHighY}{\ensuremath{9.0\,\sigma}\xspace}
\newcommand{\SigPostHighAD}{\ensuremath{13.1\,\sigma}\xspace}

\newcommand{\pythia} {\textsc{Pythia8~}}

\newcommand*{\Eqn}[1]{Eq.~#1\xspace}
\newcommand*{\Fig}[1]{Figure~#1\xspace}
\newcommand*{\Tab}[1]{Table~#1\xspace}

\journal{}

\begin{document}

\begin{frontmatter}

\title{
Enhancing the hunt for new phenomena in dijet final-states using anomaly detection filters at the High-Luminosity Large Hadron Collider
}

\author[1]{Sergei~V.~Chekanov\corref{cor1}\href{https://orcid.org/0000-0001-7314-7247}{\includegraphics[scale=0.5]{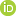}}}
\ead{chekanov@anl.gov}

\author[2]{Rui~Zhang\corref{cor1}\href{https://orcid.org/0000-0002-8265-474X}{\includegraphics[scale=0.5]{orcid_16x16.png}}}
\ead{rui.zhang@cern.ch}

\cortext[cor1]{Equal contribution}

\address[1]{HEP Division, Argonne National Laboratory,
9700 S.~Cass Avenue,
Lemont, IL 60439, USA
}

\address[2]{Department of Physics, University of Wisconsin, Madison, Wisconsin 53706, USA\\
}

\journal{ANL-HEP-183852}

\begin{abstract}
In the realm of dijet searches in high-energy physics, a significant challenge has emerged: with experiments producing more and more data, 
the traditional methods of using analytic functions to describe dijet mass spectra start to fail.
To address this, we suggest the application of an anomaly detection approach to eliminate less interesting background events based on event final states.
This method not only bypasses the limitations of conventional background models but also significantly enhances our ability to detect potential signals of new physics.
Through simulations that mimic the conditions of the upcoming High-Luminosity Large Hadron Collider, we demonstrate the strength and efficiency of this approach in dealing with large data volumes.
The integration of unsupervised machine learning into our experimental framework paves the way for a promising avenue to unveil hidden physics discoveries within the overwhelming influx of data.
\end{abstract}

\begin{keyword}
anomaly detection, unsupervised machine learning, dijet search, high energy physics
\end{keyword}

\end{frontmatter}

A search for resonances in dijet final-states is one of the primary analyses to be performed when a hadron collider reaches a new center-of-mass energy~\cite{CDF:dijet,Dzero:dijet,CDF2:dijet,Dzero2:dijet,ATLAS:dijet7a,CMS:dijet7a,ATLAS:dijet7b,CMS:dijet7b,ATLAS:dijet8,CMS:dijet8,ATLAS:dijet13,CMS:dijet13}.
Traditional dijet searches require to establish background hypotheses, which are typically empirical functions with a monotonically decreasing shape, to describe the dijet invariant-mass spectra from the Standard Model (SM) processes.
The established background hypothesis is subsequently applied to data to search for localized deviations that may indicate potential signals of heavy resonances produced by processes beyond the SM (BSM).
However, as the luminosity increases, a breakdown of the background hypotheses becomes evident~\cite{EXOT-2016-21,CMS-EXO-16-056,Chekanov_2018}.
Commonly employed analytic functions for the background description might be insufficient to describe the dijet invariant mass in high-statistics data.
Typically, this inability leads to oscillations in fit residuals, thus preventing claims about new resonances.
To address these limitations, various other techniques, such as peak finders~\cite{Chekanov:2011vj}, SWiFt~\cite{EXOT-2016-21} and functional decomposition~\cite{Edgar:2018irz} have been explored.
However, the substantial flexibility introduced by these methods may hinder the sensitivity to identify localized deviations.

In this paper, we leverage unsupervised machine learning to address this challenge. 
Instead of seeking complex functions to describe high-statistics data, we utilize an unsupervised \textit{anomaly-detection (AD) filter} in order to reduce the rate of events which are likely related to the SM processes according to their final-state topology.
Then the remaining low-statistics data can be effectively described using simple analytic functions.
As we show in this study, this approach enhances the sensitivity to detect new physics.

The advantage of our approach lies in the fact that dealing with the well-understood analytic functions studied at the Large Hadron Collider (LHC) is significantly simpler than discovering new background shapes with many free parameters.
In the case if the AD filter is not used, the possible approaches include adding additional parameters 
to the analytic fits, or using some complex numerical smoothing.
In both cases, a large flexibility of the background description leads to additional uncertainties that may ''absorb'' possible signals.

Anomaly-detection methods in high-energy physics have undergone extensive investigation, illustrating promising performance across various applications, as evidenced in Refs.~\cite{Kuusela:2011aa,DAgnolo:2018cun,Farina:2018fyg,Heimel:2018mkt,PhysRevD.99.014038,PhysRevD.99.015014,DeSimone:2018efk,Hajer:2018kqm,2019arXiv190302032R,Cerri:2018anq,Blance:2019ibf,2021JHEP153A,Dillon:2020quc,2020arXiv201103550E,2022PhRvD.105k5009B,a2022PhRvD.105e5006M,2021PhRvD.104c5003B,2022PhRvD.106e6005D,PhysRevD.107.016002}.
Dedicated endeavors have been invested in enhancing resonant bump hunting in \textsc{CWoLa}~\cite{Collins:2018epr}, \textsc{ANODE}~\cite{PhysRevD.101.075042}, \textsc{CATHODE}~\cite{Hallin:2021wme}, \textsc{SALAD}~\cite{PhysRevD.101.095004}, \textsc{FETA}~\cite{Golling:2022nkl}, \textsc{CURTAINS}~\cite{Raine:2022hht}, and many others~\cite{Aarrestad:2021oeb,kasieczka2021lhc,Karagiorgi:2022qnh}, focusing on the resonant decay products.
The utilization of unsupervised machine learning to identify ``anomalous'' events through an assessment of final-state multiplicity and kinematics has been increasingly spotlighted.
It has been argued that anomaly detection methods can help discover the vast number of event classes at the LHC, which have never been studied \cite{2020JHEP...04..030K,Chekanov:2023dby} in the past using specific BSM models.
Concrete technical implementations of this approach for model-agnostic searches demonstrated in Refs.~\cite{chekanov2021machine,Chekanov:2021pus} have been recently adopted by an ATLAS analysis~\cite{ATLAS:2023ixc}.
However, the emphasis on preserving the simplicity of the background description in the dijet final state has not received commensurate attention.

To demonstrate the power of the AD filter in this aspect, we simulate 1~$\abb$ $pp$ collision events (around 1.55 billion events) at $\sqrt{s} = 14$~TeV using the \pythia~\cite{Sjostrand:2006za, Sjostrand:2007gs} generator.
This simulation aims to emulate the statistics expected to be achieved at the High-Luminosity Large Hadron Collider~\cite{Aberle:2749422}.
The events comprise contributions from inclusive SM processes, with Quantum Chromodynamics (QCD) processes predominantly governing multijet production, while top-quark, Higgs-boson, and vector-boson productions play minor roles in comparison.
These processes are implemented in leading-order matrix elements,  with parton showering followed
by hadronization.
The NNPDF 2.3 LO~\cite{Ball:2012cx, Ball:2014uwa} parton density function, interfaced with \pythia via the LHAPDF library~\cite{Buckley:2014ana}, is used in the generation. 
For high-efficient generation of events in the region of  interest, a minimum value of transverse momentum for the matrix elements for $2\rightarrow 2$ processes is set to 1000~GeV.

To assess the potential gain for a new physics search achieved by using the AD filter, we adopt a benchmark model known as the Sequential Standard Model (SSM)~\cite{AltarelliMele}.
The SSM is an extended gauge model, which predicts the existence of additional heavy gauge bosons, commonly denoted as $W^{'}$ and $Z^{'}$.
In this study, we focus on the process of $W^{'}$ boson emission from the $s$-channel production, specifically:
\begin{equation}
qq \rightarrow W^{'} \rightarrow Z^{'} (\rightarrow q\bar{q}) 
+ W (\rightarrow \text{inclusive decay}),
\label{wzmodel}
\end{equation}
where $m_{W^{'}} = $ 3~TeV and $m_{Z^{'}} = $ 2~TeV.
The model serves as a suitable representation for a class of BSM processes that allows studies of a heavy dijet resonance ($Z^{'}$) with additional final states from the SM $W$ boson decay. 
In total, 0.1 million events were generated for the SSM process.

Electrons, muons, photons and jets are reconstructed from the stable particles with a lifetime longer than $3\cdot 10^{-10}$ seconds.
The transverse momentum of the leptons and photons must be greater than 30~GeV.
A cone energy is defined as the sum of all energies in a core of the size $0.2$ in the azimuthal angle ($\phi$) and pseudo-rapidity ($\eta$) around the true direction of the lepton or photon.
To ensure the leptons and photons are isolated, their energy must contribute more than $90\%$ of the cone energy.
The jets are reconstructed using the anti-$k_\text{T}$ algorithm~\cite{Cacciari:2008gp} with a distance parameter of $R=0.4$, implemented in the {\sc FastJet} package~\cite{Cacciari:2011ma}.
The transverse momenta of jets must be greater than $30$~GeV, and the pseudorapidity must satisfy $|\eta|<2.4$.
Jets are classified as $b$-jets if the jet momentum matches the momentum of a $b$-quark using the requirement $dR<0.4$, where $dR$ is defined as the distance in $\phi$ and $\eta$ between the direction of the jet and the $b$-quark. 
Then we require that the $b$-quark contributes more than 50\% of the total jet energy.
The $b$-jet fake rate is also included assuming an increase from 1\% 
at $p_\text{T}=30$~GeV to 6\% at $p_\text{T}=200$~GeV taken from Ref.~\cite{Aad:2015ydr}.
A detector simulation was not used.
Also, no pileup was included in the event samples.
We expect that the invariant masses above 1.2~TeV is not strongly affected by pileup.

To represent a collision event with an arbitrary number of final-states objects, we utilize the rapidity-mass matrix proposed in Ref.~\cite{CHEKANOV201992}.
This matrix effectively captures crucial details of an event, including the transverse mass and energy of each final object, as well as the invariant mass and rapidity differences between every pair of final states.
In our representation, we allow for a maximum of ten non-$b$-jets and ten $b$-jets, as well as up to five electrons, muons, and photons, along with missing transverse energy in each event.
In case an event contains fewer objects than the maximum allowed, the remaining slots in the representation are filled with zero values in the matrix.
To ensure minimal biases in the dijet mass spectrum, we exclude the dijet invariant mass variable from the leading two jets in the rapidity-mass matrix.
Other variables of the inputs cannot serve as a ``proxy'' to the dijet mass since azimuthal angles of jets, which are needed for dijet mass calculations, are not included in the inputs. 
In total, the representation results in $1295$ $(=36^2-1)$ variables for each event.

The AD filter utilizes an autoencoder implemented in \textsc{TensorFlow}~\cite{abadi2016tensorflow}.
An autoencoder is a type of neural network that learns a compressed representation of its input data in an unsupervised manner.
It consists of two key components: an encoder that reduces the input data into a lower-dimensional representation, and a decoder that reconstructs the original data from the compressed representation.
In this work, we employed a similar architecture as in Ref.~\cite{ATLAS:2023ixc}, which consists of two hidden layers with 800 and 400 neurons in the encoder and decoder, and a latent layer of 200 neurons.
The leaky ReLU~\cite{xu2015empirical} activation function is applied in all hidden and output layers.
The mean squared error between the input and reconstructed values is used as the reconstruction loss.
More sophisticated architectures are expected to achieve similar or better performance.
The autoencoder is trained on 0.7 million simulated SM events, and validated on 0.3 million  events.
The reconstructed distributions show good agreement with the distributions of the corresponding inputs in the validation dataset.
It indicates that no over-training persists and the current training dataset is sufficient for accurately capturing the intrinsic kinematics of the final states and their correlations in the SM events.

\begin{figure}[htb]
  \begin{center}
  \includegraphics[width=0.45\textwidth]{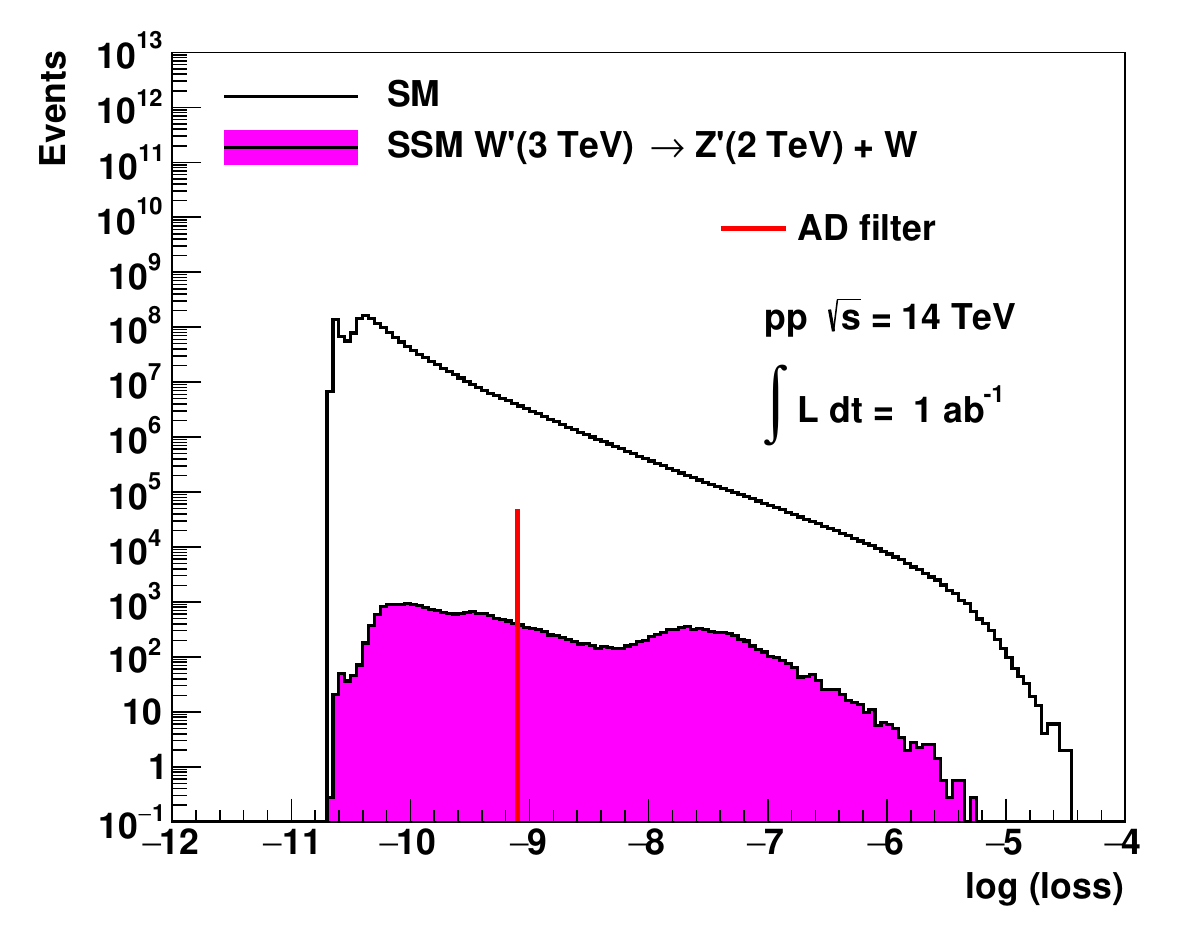}
  \end{center}
  \caption{
    Distributions of the reconstruction losses of the autoencoder for the simulated SM events and the SSM events.
    The trained AD filter rejects events with $\logloss < -9.1$.
 }
 \label{fig:loss}
\end{figure}

The trained AD filter is then applied to the remaining events.
The reconstruction loss of the autoencoder is shown in \Fig{\ref{fig:loss}}.
Events with small reconstruction losses ($\logloss < -9.1$), indicated with the vertical line in \Fig{\ref{fig:loss}}, are excluded, rejecting 98\% of the SM events while keeping 38\% of the benchmark SSM events.
This criterion is used in Ref.~\cite{ATLAS:2023ixc} although it can be further optimized for other analyses.
Two distinct scenarios of different statistics are investigated: \lumismall and \lumilarge.
In addition to using the AD filter, we also investigate an alternative approach by applying a requirement on rapidity difference $|y^{*}| = |(y^{j_1} - y^{j_2}) / 2| < 0.6$ between the two leading jets, $j_1$ and $j_2$, which is commonly used in traditional dijet searches (e.g.\ Refs.~\cite{ATLAS:dijet13,CMS:dijet13}) to suppress the $t$-channel QCD background.
This selection removes 45\% of the SM events in our study. 

The dijet invariant mass, $\mjj$, is computed using the two leading jets with the highest transverse momenta, $p_\text{T}$, in the event.
In the following analysis, we apply \Eqn{\ref{eq:bkg}} to fit the $\mjj$ spectrum:
\begin{equation}
  \text{5 par}: f(x) = p_1 (1 - x)^{p_2} x^{p_3 + p_4\ln x+ p_5\ln^2 x},
\label{eq:bkg}
\end{equation}
where $x = \mjj /\sqrt{s}$ and $p_i$\ represent five free parameters.
Similar expressions show excellent performance in previous dijet searches~\cite{CDF2:dijet,ATLAS:dijet7a,CMS:dijet7a,ATLAS:dijet7b,CMS:dijet7b,ATLAS:dijet8,CMS:dijet8,ATLAS:dijet13,CMS:dijet13}.
In this study, the fit begins at 1200~GeV to  focus on the higher-mass region of interest.

\begin{figure*}[htb]
    \begin{center}
    \subfloat[\lumismall]{\includegraphics[width=0.4\textwidth]{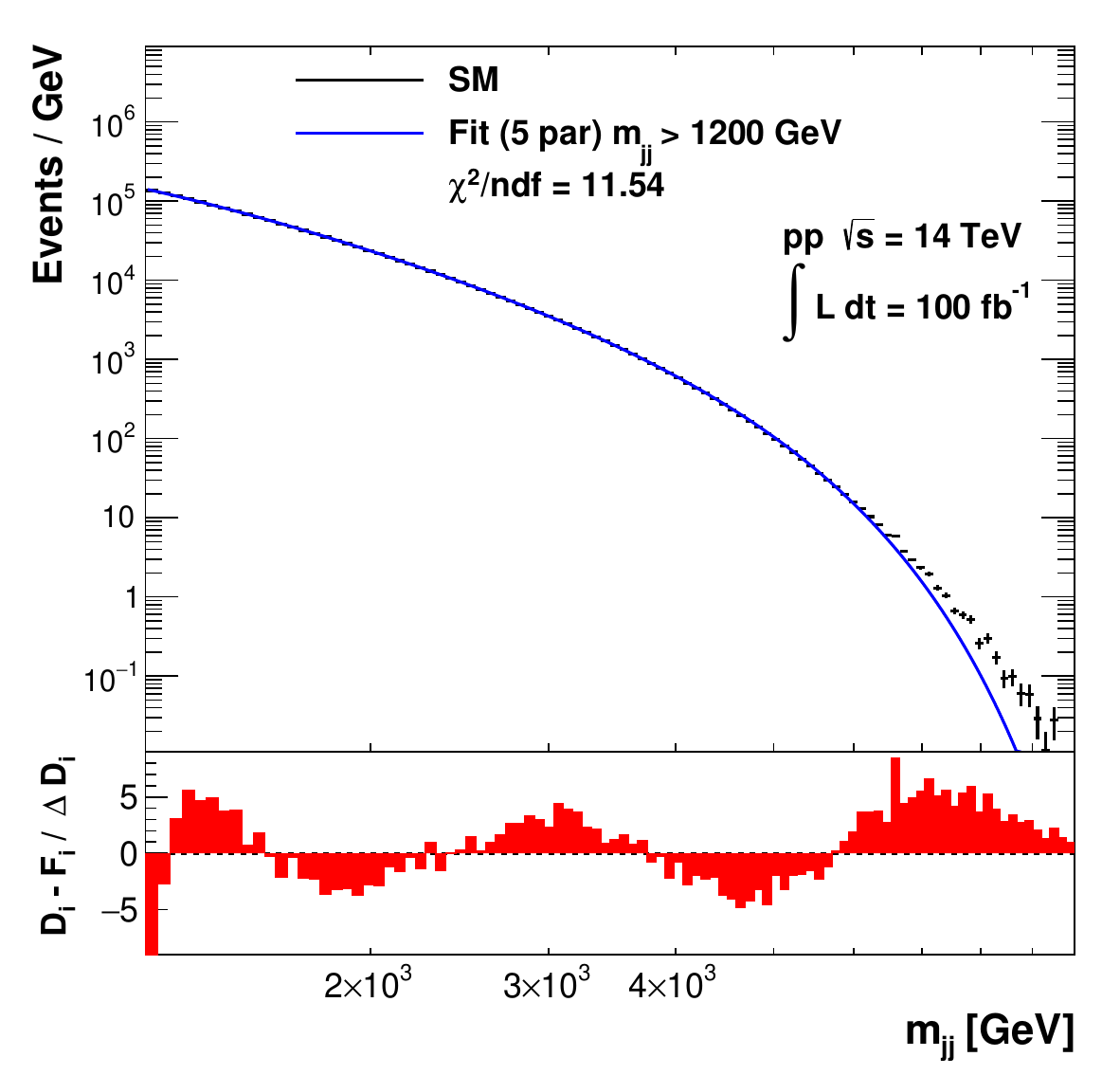}}
    \subfloat[\lumilarge]{
    \includegraphics[width=0.4\textwidth]{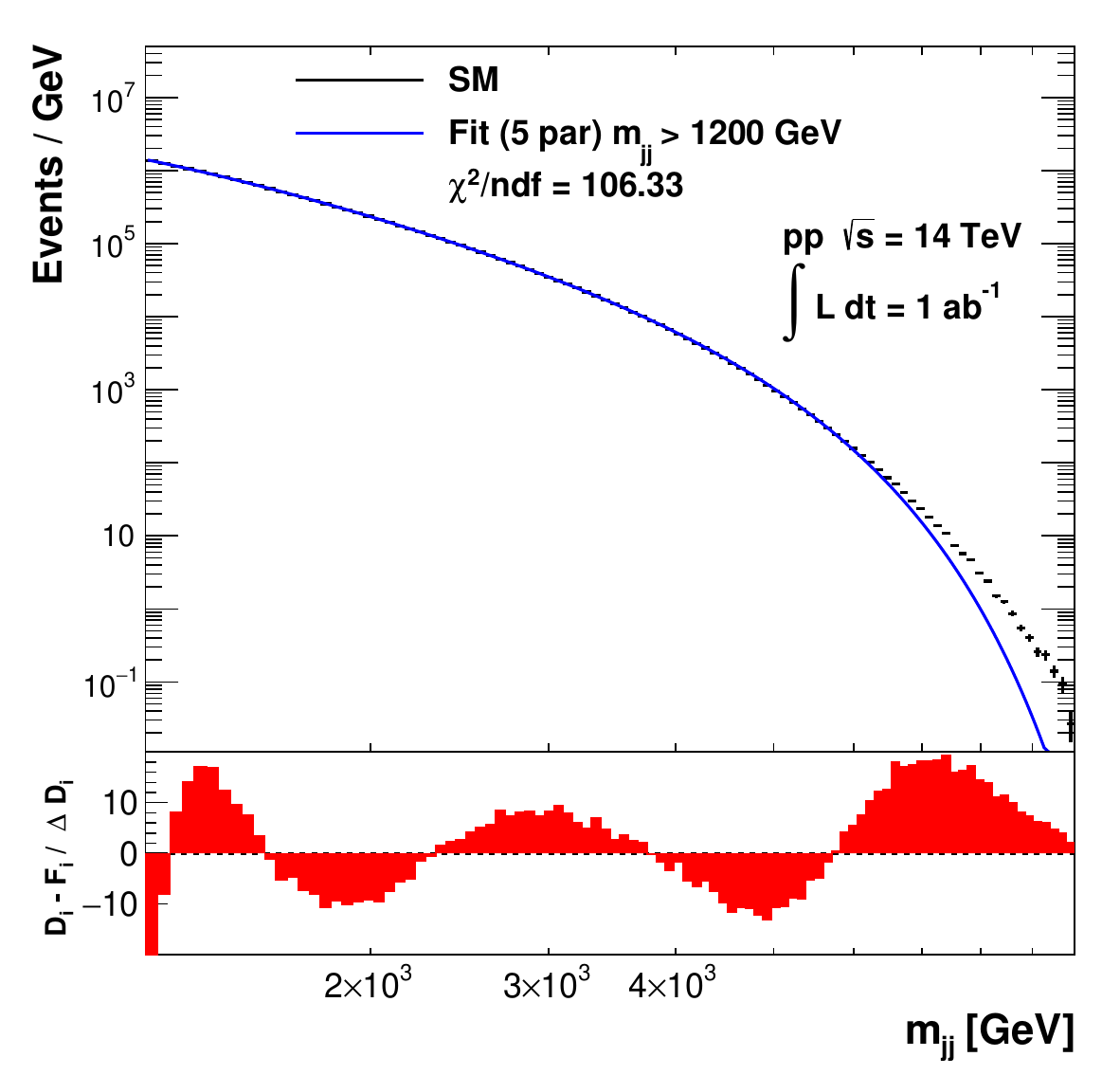}
    }
  \end{center}
  \caption{
    The distributions of the dijet invariant mass of the simulated SM events for the \lumismall and \lumilarge scenarios and functional fits using \Eqn{\ref{eq:bkg}}.
    The bottom panels show the significances of deviations of the fit.
    A ``wave'' pattern for the deviations is observed indicating failures of the fit to describe the simulated SM events.
  }
  \label{fig:mass_jj_beforecut}
\end{figure*}

\Fig{\ref{fig:mass_jj_beforecut}} demonstrates that \Eqn{\ref{eq:bkg}}, after the $\chi^2$ minimization, fails to describe the data prior to applying either the $\ycut$ selection or the AD filter.
The bottom panels show the deviations of the fit value ($F_i$) from the simulated data ($D_i$) in the $i$-th bin, scaled by $1/\Delta D_i$, where $\Delta D_i$ is the statistical uncertainty of the data in that bin.
The observed ``wave'' pattern for the deviations is a distinctive trait of this class of functions when they are unable to describe the data.
The fit becomes reasonable when the data sample is 
reduced to 1~fb$^{-1}$, as demonstrated in \Fig{\ref{fig:mass_jj_beforecut_1fb}} in \ref{A1}.
As statistics increase, the failure becomes more pronounced, indicating the limitation of the function in representing the high-statistics data.
Adding an extra free parameter to this function does not solve this problem as shown in \Fig{~\ref{fig:mass_jj_beforecut_p6}} in \ref{A1}.
Searching for a new physics under this situation, assuming relatively small contributions from heavy resonances,
becomes impossible.

To understand the discovery potential of the simulated data to new resonance phenomena, we inject simulated SSM events into the SM events.
The injected amount is such that the following quantity, defined as a measure of the significance~\cite{cowan2012}, is approximately 3:
\begin{equation}
  Z_A = \sqrt{2 \left(  (s+b)\ln \left( 1+\frac{s}{b} \right) -s \right)}
\label{eq:ZA}
\end{equation}
Here $s$ is the injected number of the SSM events and $b$ is the number of SM events.
This injection corresponds to approximately 12 thousand SSM events injected into approximately 17 million SM events in the region $\mjj>1200$~GeV after applying the \ycut selection. 

We also define two significances as follows:
\begin{itemize}
  \item The pre-fit significance:
    \begin{equation}
      Z_A^{\text{pre}} = \sqrt{\sum_i^{\text{bins}} Z_i^2}, \;\; Z_i = \sqrt{2 \left(  (s_i+b_i)\ln \left( 1+\frac{s_i}{b_i} \right) -s_i \right)},
      \label{eq:prefitsig}
    \end{equation}
    where $s_i$ and $b_i$ are the number of SSM events and the number of SM events in the $i$-th bin before fit, respectively.
    The pre-fit significance is expected to exceed the ``injected'' significance calculated from the event yields since it is primarily influenced by the signal-over-background ratio around the peak position.
  \item The post-fit significance ($Z_A^{\text{post}}$) is defined according to \Eqn{\ref{eq:ZA}} with $s=\sum (D_i-F_i)$ and $b=\sum F_i$.
  The summations run from the peak position, defined by the maximum value of $(D_i-F_i)/\Delta D_i$, to the left and the right side of the peak, and stop when this quantity becomes negative, or the bin center is outside of the resonance width, which is $10\%$ of the mass.
  The fit values $F_i$ are obtained through the fitting process to the SM events and injected SSM events using \Eqn{\ref{eq:bkg}}, while $D_i$ represents the prediction from both sets of events.
\end{itemize}
The post-fit significance is usually smaller than the pre-fit significance because the fit with only the background function to data could ``absorb'' some injected signals.
Nonetheless, it provides a reliable assessment of the observed deviations from the background hypothesis and helps identify potential new physics phenomena.
The post-fit also shows a broad agreement with the local significance employed in \textsc{BumpHunter}~\cite{bumphunter}, which is a widely adopted algorithm in hadron collider experiments.

\begin{figure*}[htb]
    \begin{center}
        \subfloat[SM events with \ycut]{\includegraphics[width=0.4\textwidth]{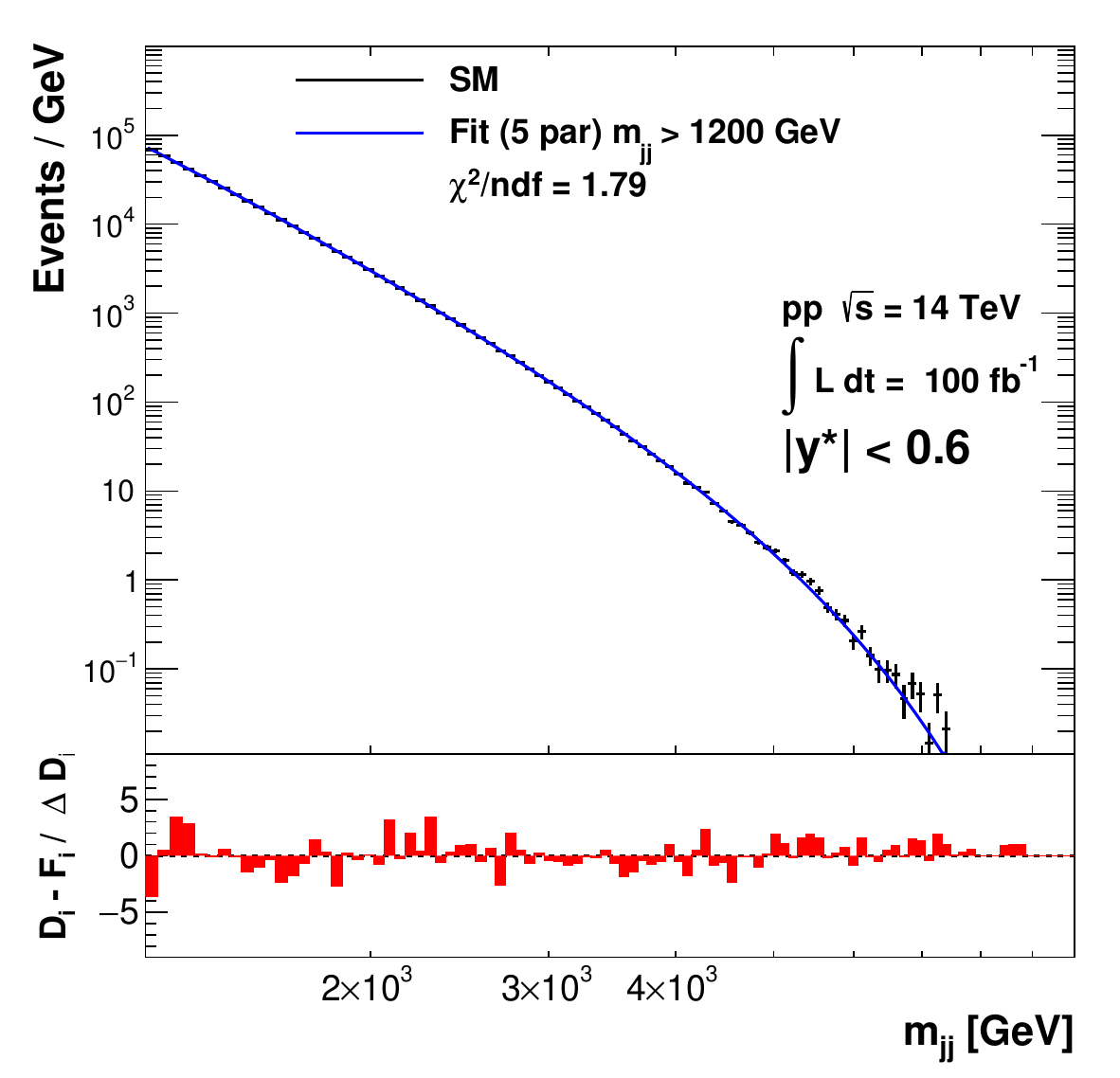}}
        \subfloat[SM events with AD filter]{\includegraphics[width=0.4\textwidth]{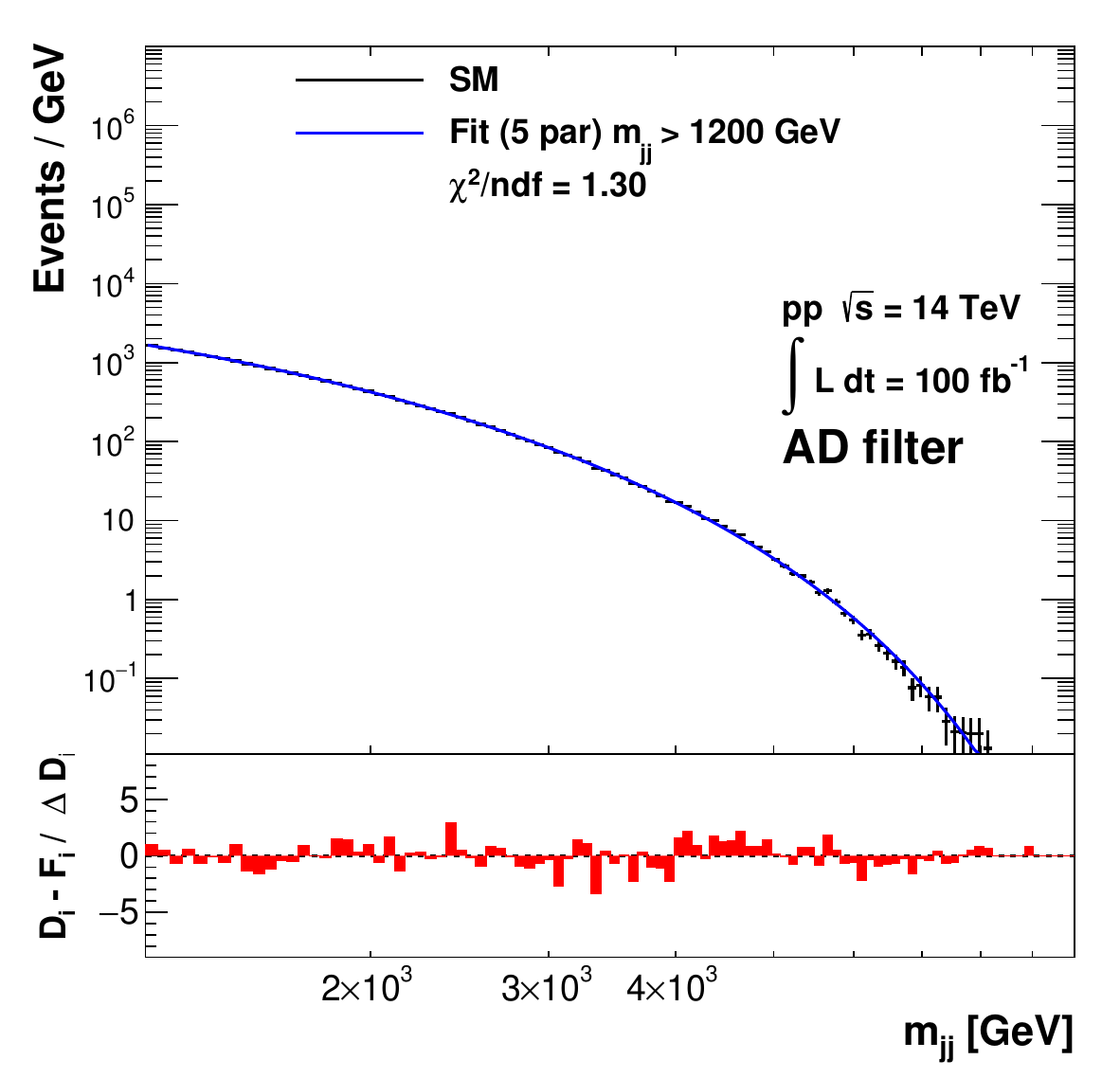}} \\
        \subfloat[SM and SSM events with \ycut]{\includegraphics[width=0.4\textwidth]{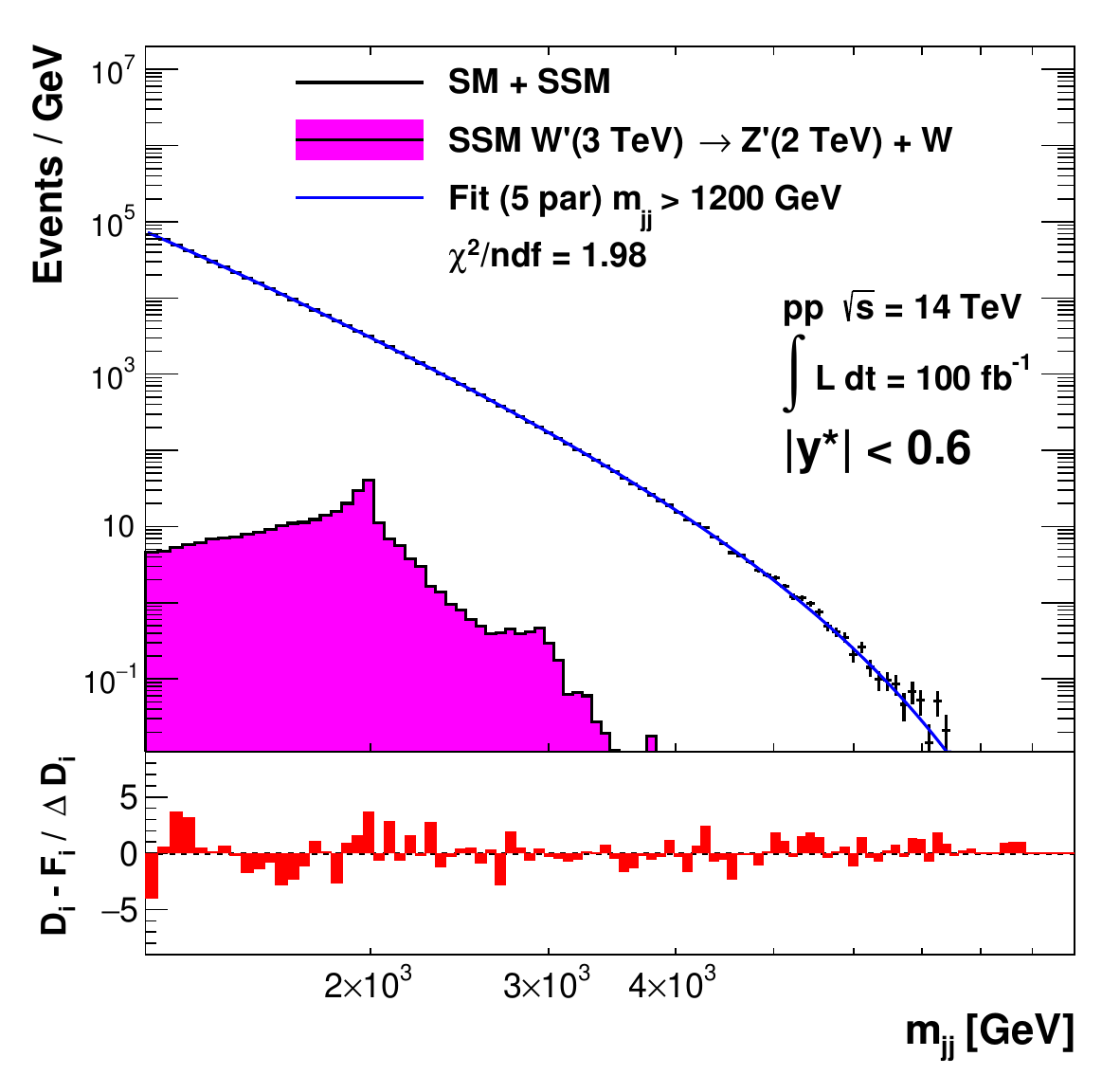}}
        \subfloat[SM and SSM events with AD filter]{\includegraphics[width=0.4\textwidth]{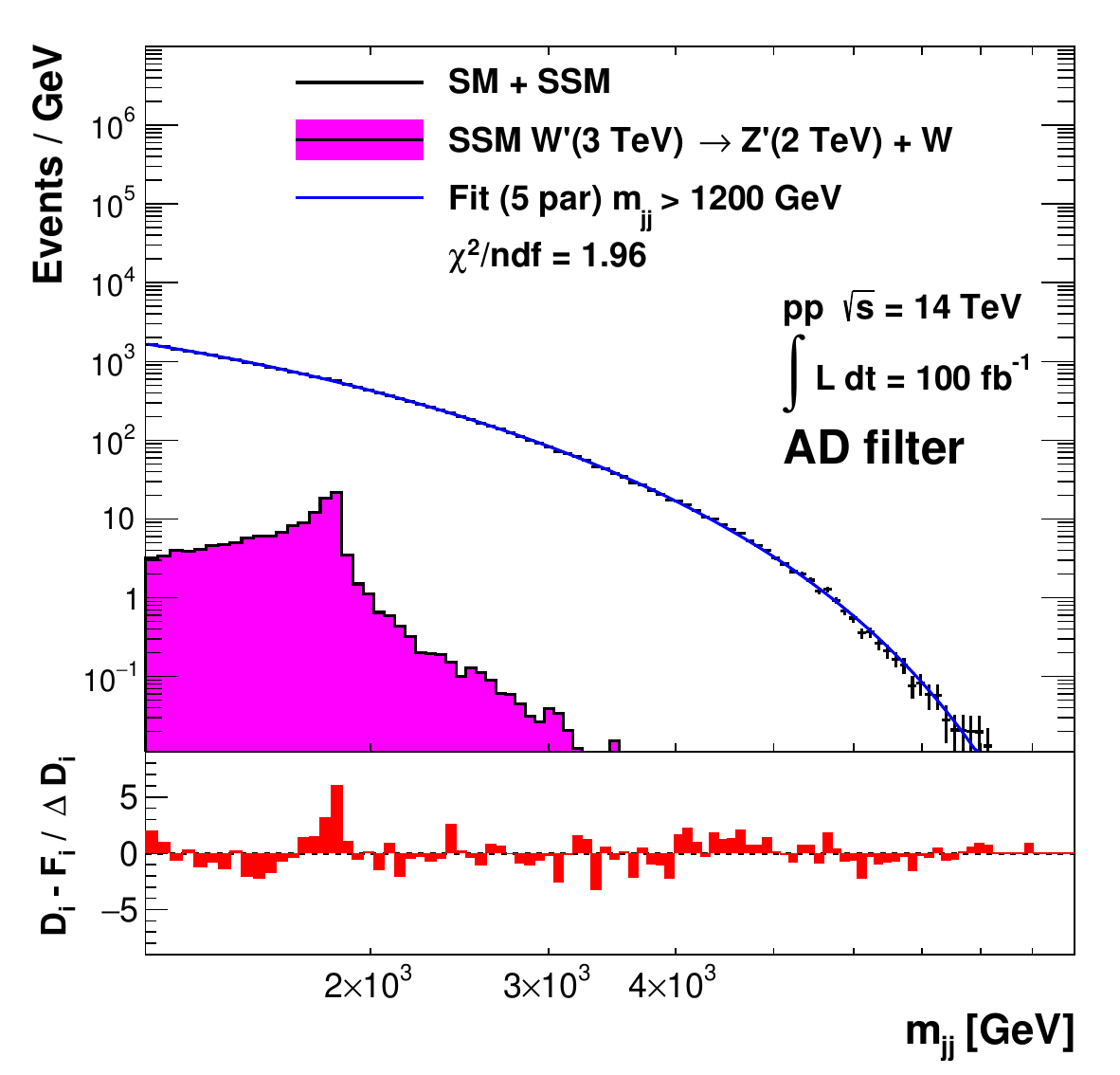}}
  \end{center}
  \caption{
   The dijet invariant mass distributions for the simulated SM events (a) (b) and with injected SSM events (c) (d) in the \lumismall scenario.
   The two selections of \ycut or the AD filter are applied, respectively. 
  }
  \label{fig:mass_jj_100fb}
\end{figure*}

After applying either the selection on the rapidity difference \ycut or the AD filter, we observe a significant improvement in the fit quality.
\Fig{\ref{fig:mass_jj_100fb}} illustrates the fit to the $\mjj$ distribution in the \lumismall scenario, both with and without SSM events injected.
Notably, both selections exhibit smaller $\chi^2$/ndof (number of degrees of freedom) compared to the fit shown in \Fig{\ref{fig:mass_jj_beforecut}}.
However, it is evident that the fit quality achieved with the AD filter surpasses that of the \ycut selection.
The post-fit significance of the benchmark signal with the AD filter is measured as \SigPostLowAD, marking a substantial improvement from the \SigPostLowY achieved with the \ycut selection.

\begin{figure*}[htb]
    \begin{center}
      \subfloat[SM events with \ycut]{\includegraphics[width=0.4\textwidth]{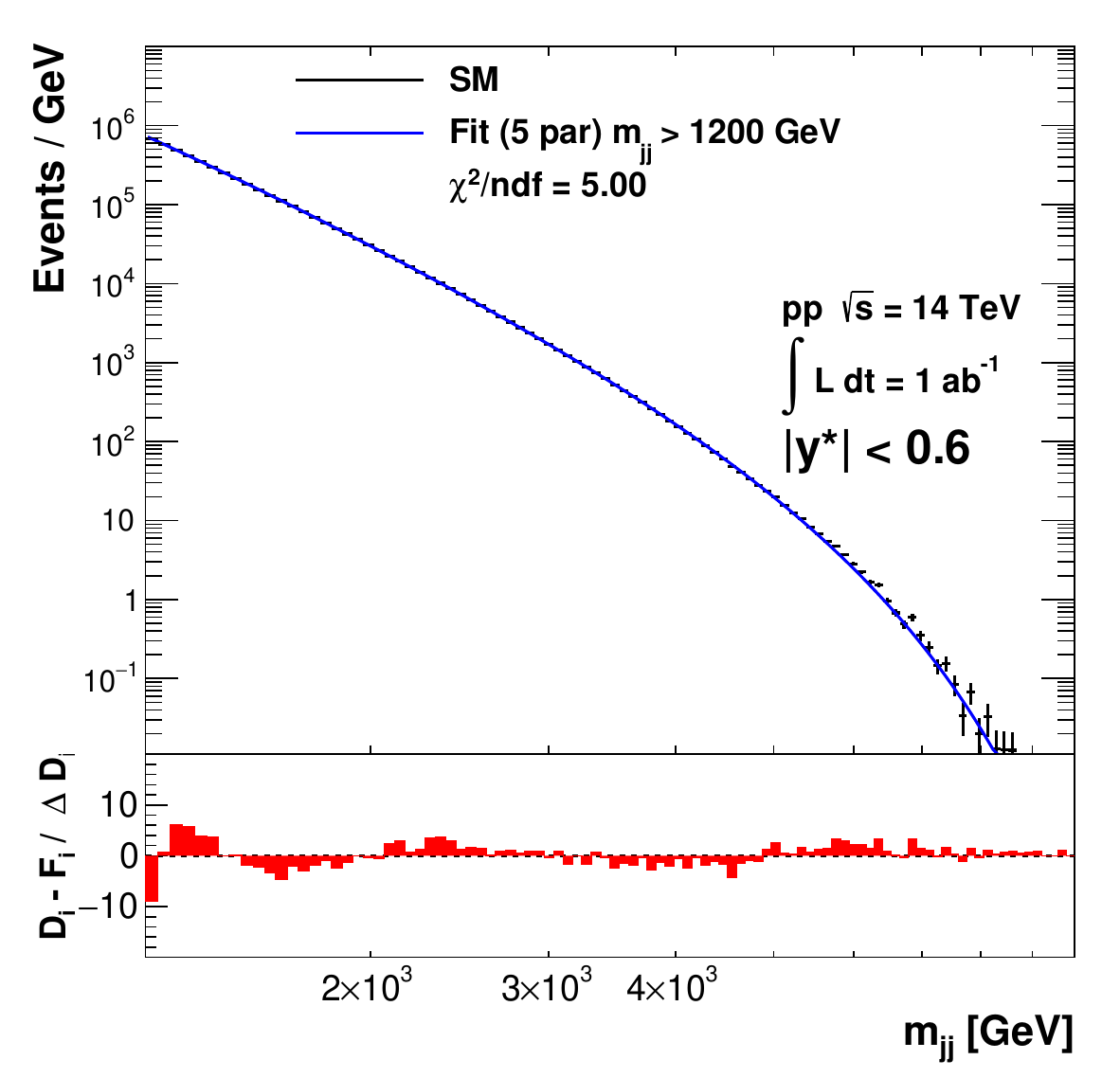}}
      \subfloat[SM events with AD filter]{\includegraphics[width=0.4\textwidth]{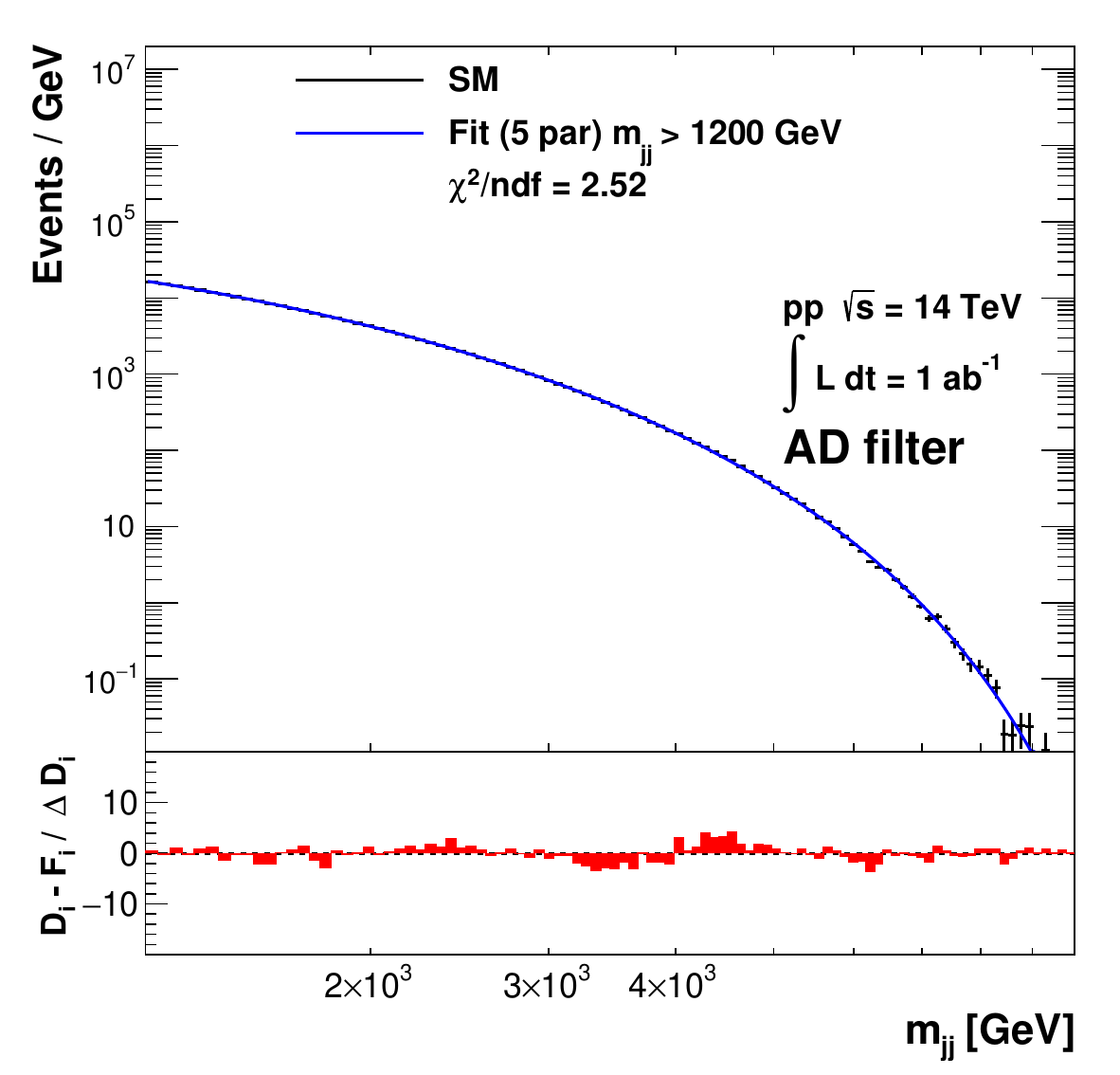}} \\
      \subfloat[SM and SSM events with \ycut\label{fig:mass_jj_1ab_c}]{\includegraphics[width=0.4\textwidth]{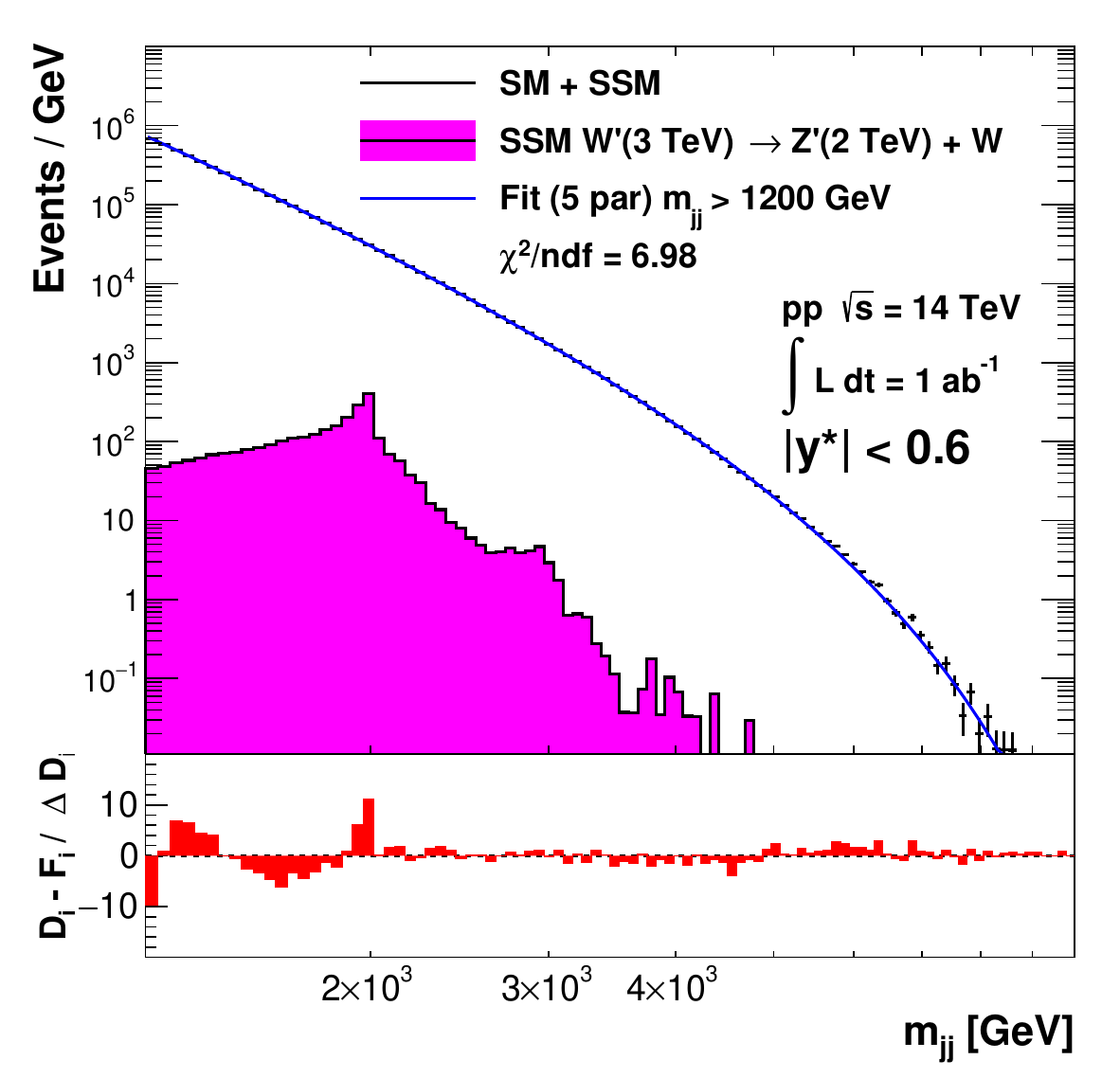}}
      \subfloat[SM and SSM events with AD filter]{\includegraphics[width=0.4\textwidth]{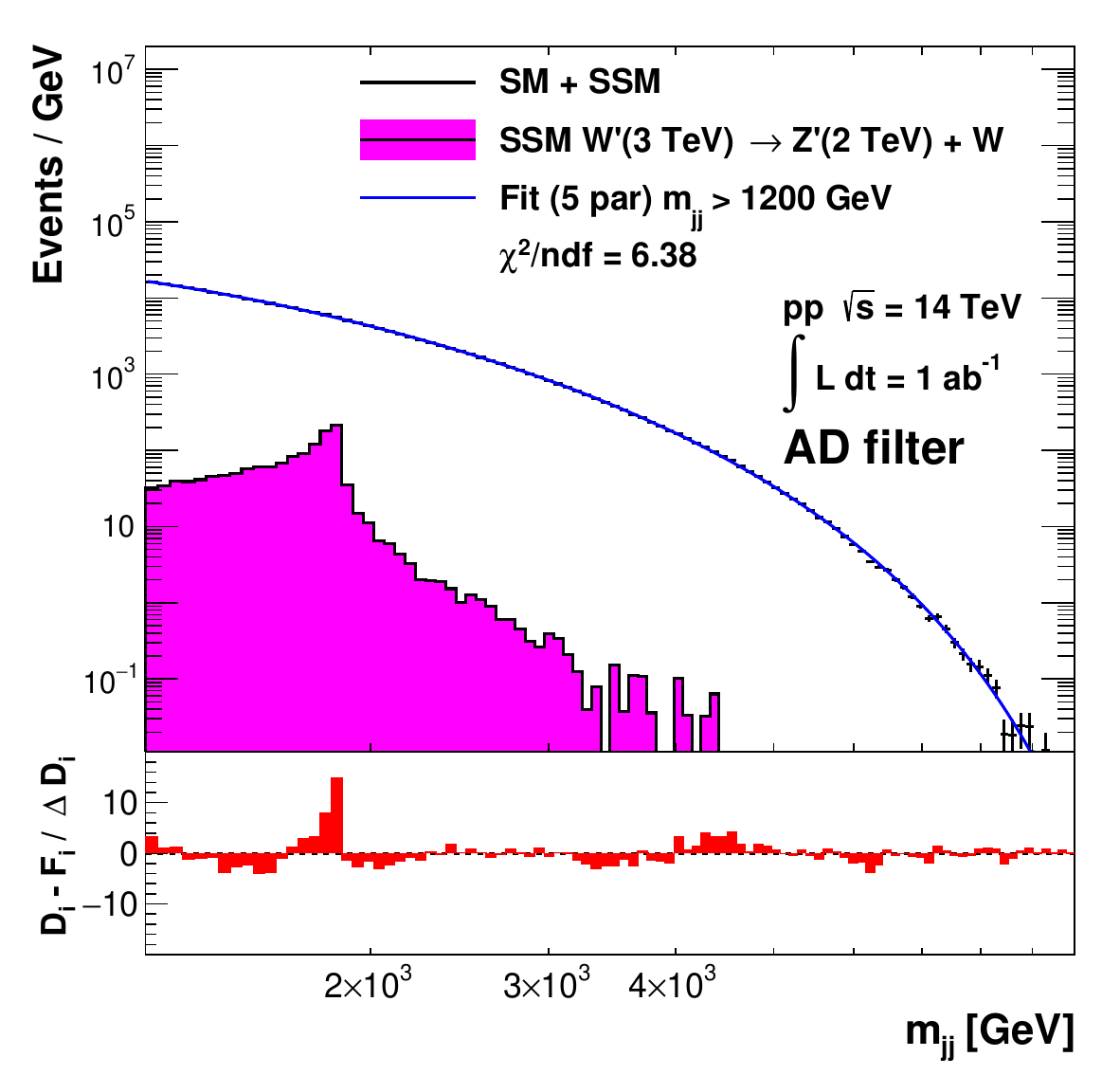}}
  \end{center}
  \caption{
   The dijet invariant mass distributions for the simulated SM events (a) (b) and with injected SSM events (c) (d) in the \lumilarge scenario.
   The two selections of \ycut or the AD filter are applied, respectively. 
  }
  \label{fig:mass_jj_1ab}
\end{figure*}

\Fig{\ref{fig:mass_jj_1ab}} presents the fit results in the \lumilarge scenario, revealing a more pronounced enhancement with the AD filter compared to the \ycut selection.
Evidently, the AD filter not only improves the fit quality but also yields an impressive increase in the post-fit significance, rising from \SigPostHighY to \SigPostHighAD, further affirming the superior performance of the AD filter in this scenario.
In both luminosity scenarios, the improvement in the pre-fit significance suggests that the AD filter outperforms the \ycut selection in terms of background rejection. 
\Tab{\ref{tab:sig}} provides a summary of the pre- and post-fit significances for both selections under both scenarios, highlighting the clear advantage of the AD filter over the \ycut selection in terms of signal significance improvement.

\begin{table*}[htb]
    \centering
    \begin{tabular}{c|cccccc}
        \toprule
        \multirow{2}{*}{\makecell{Integrated\\luminosity}} & \multicolumn{2}{c}{\makecell{Pre-fit\\significance}} & \multicolumn{2}{c}{\makecell{Post-fit\\significance}} & \multicolumn{2}{c}{\makecell{Fit $\chi^2$/ndf\\SM events}} \\
         & \ycut & AD filter & \ycut & AD filter & \ycut & AD filter \\
        \midrule
        \lumismall & \SigPreLowY  & \SigPreLowAD  & \SigPostLowY  & \SigPostLowAD  & 1.8 & 1.3 \\
        \lumilarge & \SigPreHighY & \SigPreHighAD & \SigPostHighY & \SigPostHighAD & 5.0 & 2.5 \\
        \bottomrule
    \end{tabular}
    \caption{
        Pre- and post-fit significances for the injected benchmark SSM signal under the \lumismall and \lumilarge scenarios. The last two columns show the $\chi^2/$ndof values for the background fits with \Eqn{\ref{eq:bkg}}.
        The \ycut and the AD filter selections are compared.
    }
  \label{tab:sig}
\end{table*}

\FloatBarrier

\begin{figure*}[htb]
    \begin{center}
        {\includegraphics[width=0.5\textwidth]{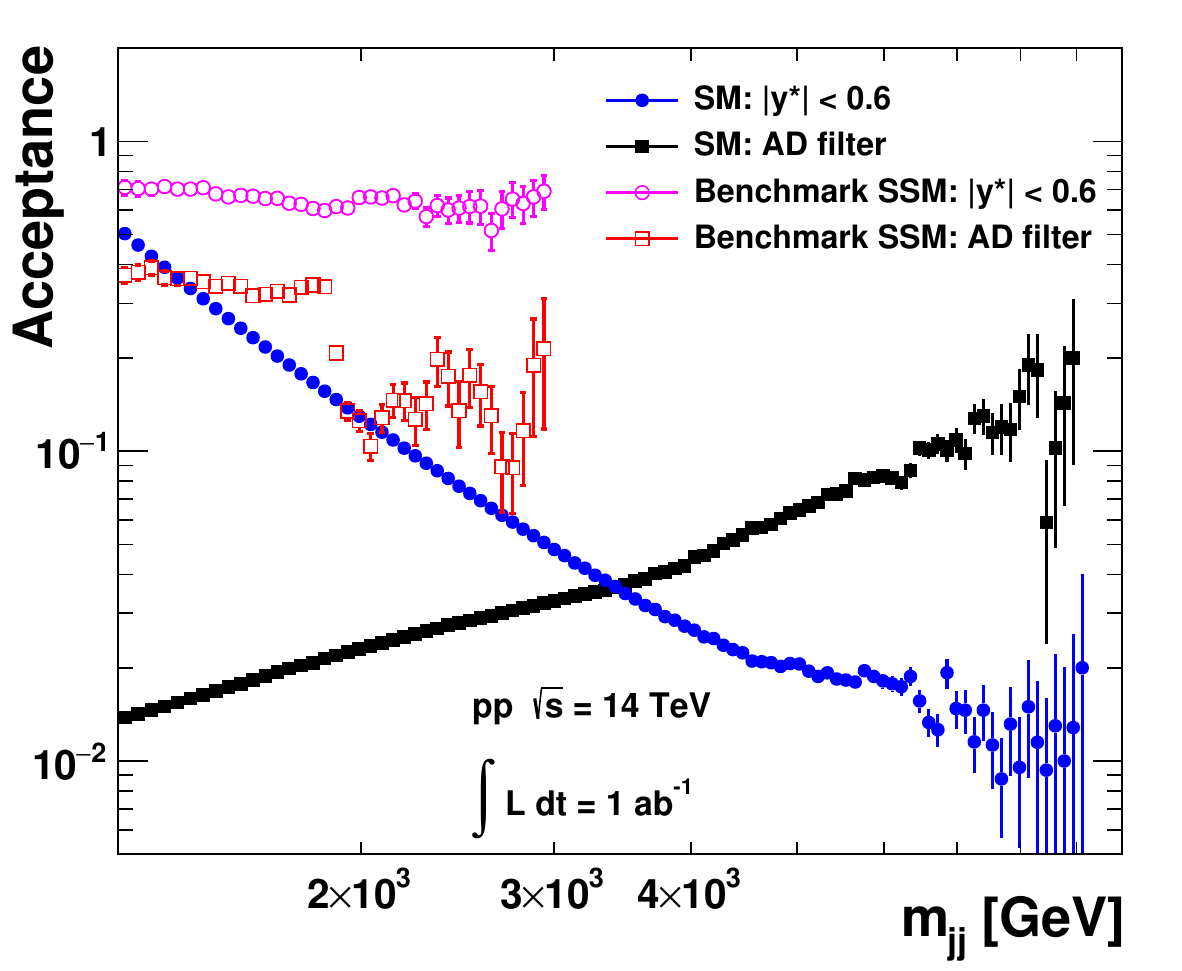}}
    \end{center}
    \caption{
        The acceptance of the \ycut selection or the AD filter, evaluated for both the SM events or the benchmark SSM events.
        Acceptance is determined as the ratio of events retained after applying either the \ycut criterion or the AD filter to the total number of events within each specific mass bin.
    }
    \label{fig:mass_jj_acceptance}
\end{figure*}

The acceptance of the AD filter is depicted for both SM events and the benchmark SSM events in \Fig{\ref{fig:mass_jj_acceptance}}.
The figure also presents a comparison of the acceptance achieved using the \ycut selection.
The acceptances are determined by computing the ratio of the number of events within each bin after the selections, relative to the number of events before the selections.
For the SM events, the AD filter exhibits a significantly lower acceptance for the low dijet mass region, gradually rising as $\mjj$ increases.
This pattern effectively retains high mass events, which are more intriguing, while concurrently suppressing low mass events that are more likely attributable to the SM process.
This behavior aligns with our expectations, as a larger invariant mass corresponds to higher values of the inputs to the autoencoder, consequently resulting in a larger reconstruction loss.
The acceptance pattern for the $\ycut$ selection contrasts significantly with that of the AD filter.
While the $\ycut$ selection shows low acceptance at large dijet masses for the SM events, its effectiveness is less pronounced for low dijets masses.
This phenomenon arises from the increased contribution of the $t$-channel QCD process.

The acceptance for the benchmark SSM events remains relatively flat when considering the \ycut criterion, whereas a distinct non-linear pattern emerges for the benchmark SSM events utilizing the AD filter.
Within the AD filter scenario, a reduction beyond 2~TeV is observed, which is driven by underlying kinematics.
When the combined energy of the two leading jets exceeds 2~TeV, the available phase-space for the production of additional jets in the SSM events is reduced.
Consequently, events exhibit a diminished final-state multiplicity, thereby contributing to a lower reconstruction loss. This effect is attributed to the fact that a small number of non-zero cells in the rapidity-mass matrix typically leads to small loss values. 
The acceptance experiences a gradual increase due to the influence of the energy hardness of the two leading jets.
This is another feature of the inputs where objects with large energies lead to larger values of the cells of the rapidity-mass matrix and, consequently, to an increase in the loss values.
Another dip in acceptance becomes evident when the dijet invariant mass reaches 3~TeV, aligning with the disappearance of the broader bump around this mass that is otherwise shown in Figure~\ref{fig:mass_jj_1ab_c}.
This bump is likely a result of the decay products from the $Z^{'}$ and $W$ bosons merging into two jets, and consequently, the suppression of final state multiplicity is enhanced.

In addition, we also studied two further scenarios to cross check the observed behaviour in the benchmark SSM events.
In the first scenario where only $W\rightarrow l\nu$ decays are mandated, the signal acceptance of the AD filter undergoes a notable increase by a factor of 2 to 3, reaching an efficiency of 0.8, illustrated in 
Figure~\ref{fig:mass_jj_acceptance_3tev_decays}(a) in \ref{A2}.
This effect is due to the fact that the rapidity-mass matrix elements for leptons and transverse masses are more frequently activated, making
leptonic decays relatively more anomalous compared to the case with  $W$ inclusive decay. 
Consequently, SSM events with the hadronic decays of $W$ are less anomalous. 
The acceptance for the SSM events with the hadronic decays of $W$ is lower than for the leptonic decays, as shown in \Fig{~\ref{fig:mass_jj_acceptance_3tev_decays}}(b) in \ref{A2}. 

In the second scenario, where the $W^{'}$ mass is elevated to 5~TeV within the benchmark SSM, a distinct trend emerges, as shown in Figure~\ref{fig:mass_jj_acceptance_5tev} in \ref{A2}.
The signal acceptance of the AD filter increases to about 0.8--0.9.
Moreover, the larger $W^{'}$ boson mass allows greater phase-space for higher final state multiplicities, which leads to events that are less aligned with the majority of the SM events.
However, since in this high-mass region, the \ycut becomes more effective in rejecting the SM background than the AD selection, there appears to be no advantage in using the AD in the ``tail'' of the dijet mass spectra for the SSM signal.
These observations, however, may depend on the kinematic features of the BSM models in the high-mass regime.

It is also essential to emphasize that the observed improvements in the discovery potential of heavy particles depends on the selection criterion applied to the loss value.
In this study, we employed the requirement $\logloss > -9.1$ inspired by the ATLAS analysis~\cite{ATLAS:2023ixc}.
However, certain BSM models yielding low missing transverse energy or lepton multiplicities, such as excited-quark $q^*$ models~\cite{Baur:1987ga}, may result in a reduction of the reconstruction loss, making this requirement sub-optimal. Nevertheless, since the rapidity difference is an integral component of the rapidity-mass matrix, we anticipate the AD filter to maintain its superior performance over the \ycut selection following an appropriate adjustment of the loss cut.

Note that if the goal of the anomaly detection technique is to find unusual events in a ``model-agnostic'' approach, the decision on the value of the loss cut should be determined from the data themselves, i.e.\ without using BSM models.
This topic of defining the most appropriate anomaly region without using BSM models is beyond the scope of this paper.

In summary, with the continuous expansion of experimental data in terms of statistics, conventional techniques for characterizing dijet invariant mass distributions are encountering inherent challenges.
In this context, the application of unsupervised methods, such as the anomaly-detection filter discussed in this paper, emerges as an appealing strategy to reduce the ambiguity associated with finding complex fit functions with many free parameters. 
By selectively identifying and suppressing background events based on their final-state signatures, this approach effectively tackles the shortcomings associated with empirical background models.
As we have shown through the simulations, the use of the AD filter results in strong observations of signals in the scenarios where conventional techniques can only offer weak evidence or even fail to draw any conclusion in the absence of a $|y^{*}|$ cut.
This potential extends beyond the benchmark model, as the topology of the BSM model used in the study is quite general.
It is worth noting that the acceptance of the AD filter is influenced not only by the rapidity difference between the two leading jets but also by the overall configuration of the final states within BSM scenarios.
Our findings suggest that, especially in the era of high statistics, the integration of unsupervised techniques offers a promising path to advancing the accuracy and power of anomaly searches in the dijet mass spectrum.

\section*{Acknowledgments}
The submitted manuscript has been created by UChicago Argonne, LLC, Operator of Argonne National Laboratory (“Argonne”). Argonne, a U.S. 
Department of Energy Office of Science laboratory, is operated under Contract No. DE-AC02-06CH11357. The U.S. Government retains for itself, 
and others acting on its behalf, a paid-up nonexclusive, irrevocable worldwide license in said article to reproduce, prepare derivative works, 
distribute copies to the public, and perform publicly and display publicly, by or on behalf of the Government.
The Department of Energy will provide public access to these results of federally sponsored research in accordance with the 
DOE Public Access Plan. \url{http://energy.gov/downloads/doe-public-access-plan}. Argonne National Laboratory’s work was 
funded by the U.S. Department of Energy, Office of High Energy Physics (DOE OHEP) under contract DE-AC02-06CH11357. The Askaryan Calorimeter Experiment was supported by the US DOE OHEP under Award Numbers DE-SC0009937, DE-SC0010504, and DE-AC02-76SF0051.
RZ is supported by DE-SC0017647.

\section*{Data Availability}
The data and machine learning model used to support the findings of this study can be accessed via zenodo.org~\cite{zhang_rui_2023_8219104}.

\section*{References}
\bibliographystyle{elsarticle-num}
\def\bibname{\Large\bf References}
\bibliography{references}

\newpage
\appendix
\setcounter{figure}{0}
\renewcommand\thefigure{A.\arabic{figure}}
\addcontentsline{toc}{section}{Appendix}

\section{Fit examples}
\label{A1}

\begin{figure*}[htb]
    \begin{center}
    \subfloat[\lumismall]{\includegraphics[width=0.37\textwidth]{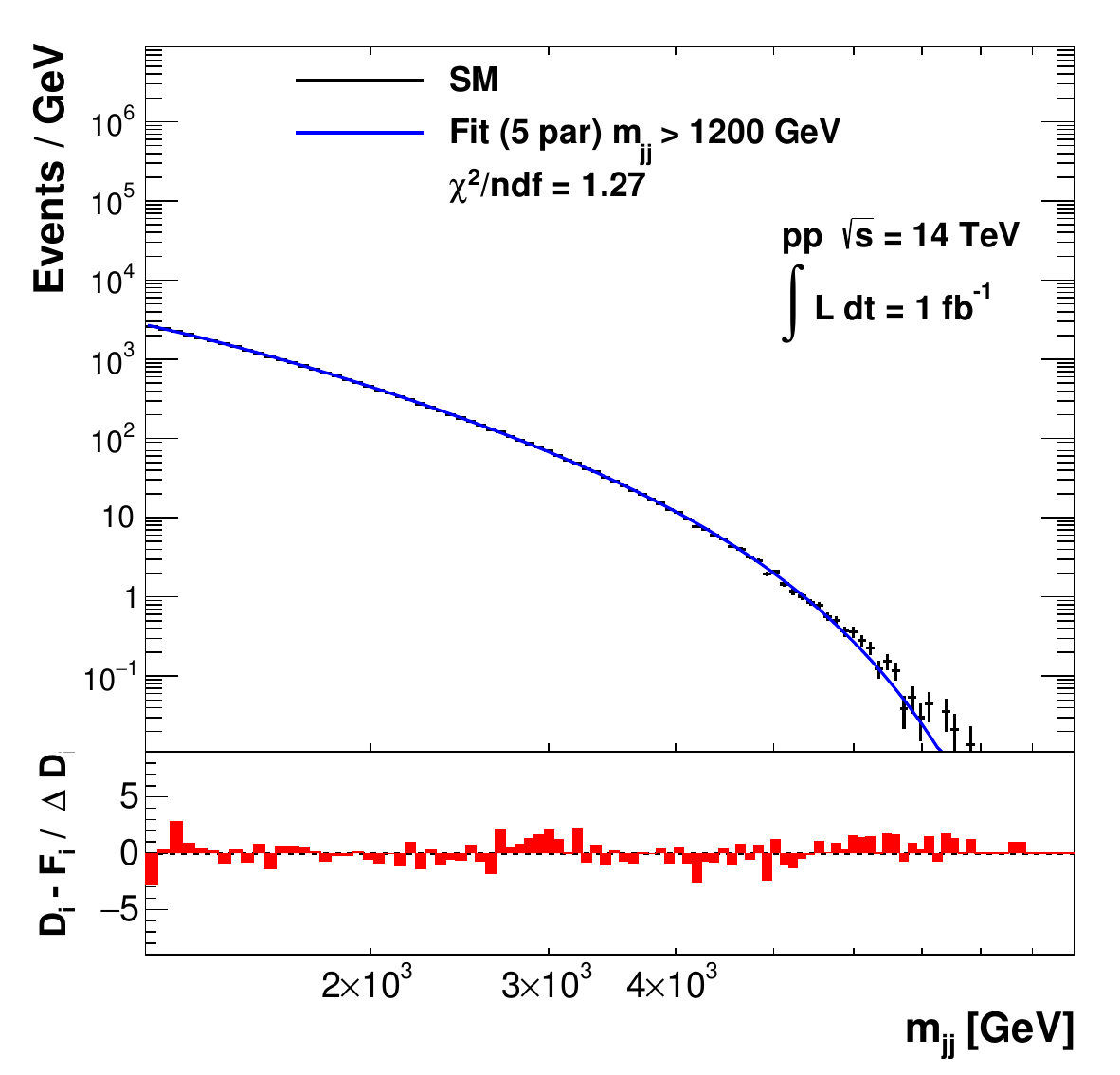}}
  \end{center}
  \caption{
    The distributions of the dijet invariant mass of the simulated SM events for the 1~fb$^{1}$ scenario and functional fits using \Eqn{\ref{eq:bkg}}.
    The bottom panels show the significances of deviations of the fit. 
  }
  \label{fig:mass_jj_beforecut_1fb}
\end{figure*}

\begin{figure*}[htb]
    \begin{center}
    \subfloat[\lumismall]{\includegraphics[width=0.37\textwidth]{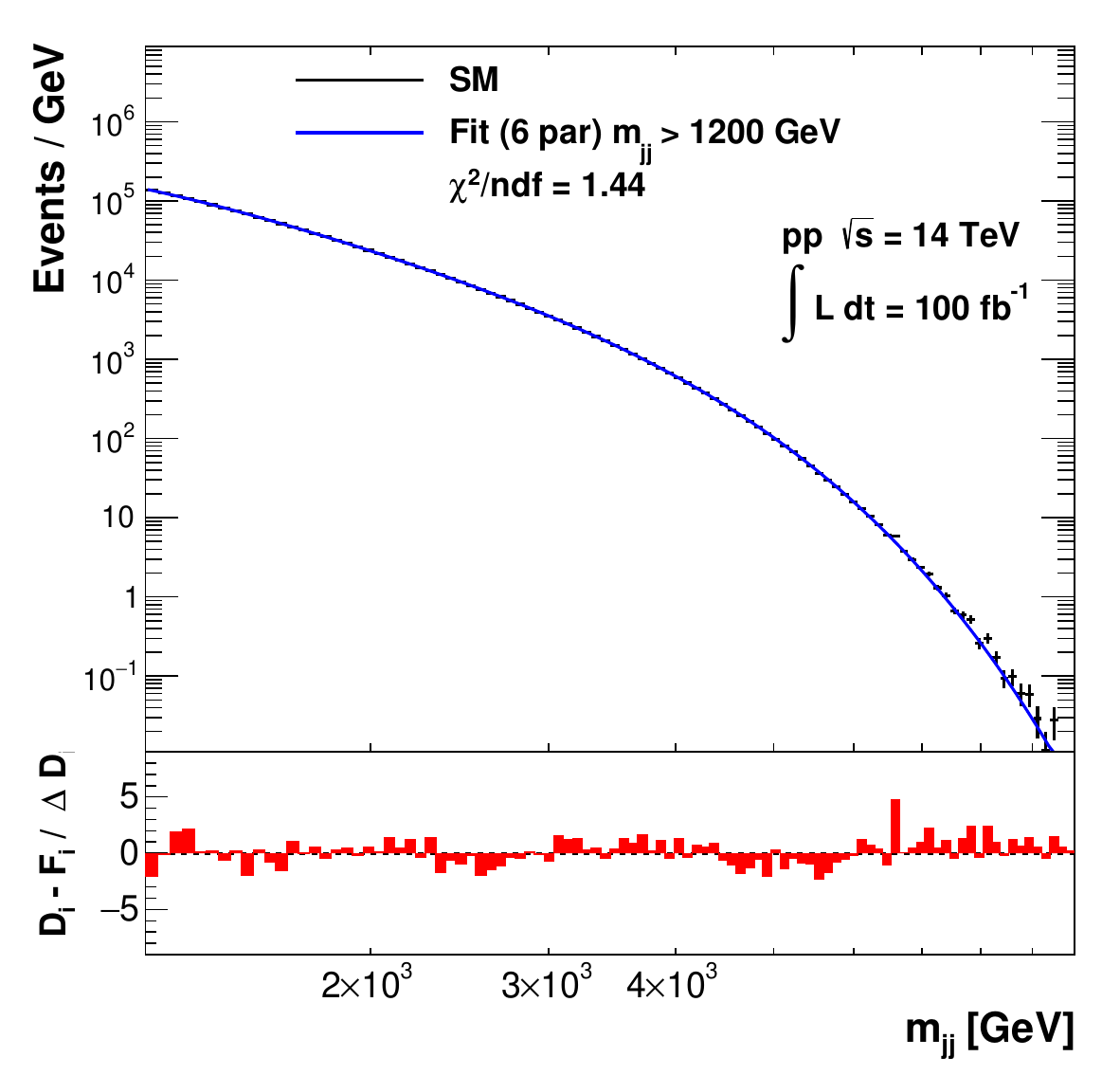}}
    \subfloat[\lumilarge]{
    \includegraphics[width=0.4\textwidth]{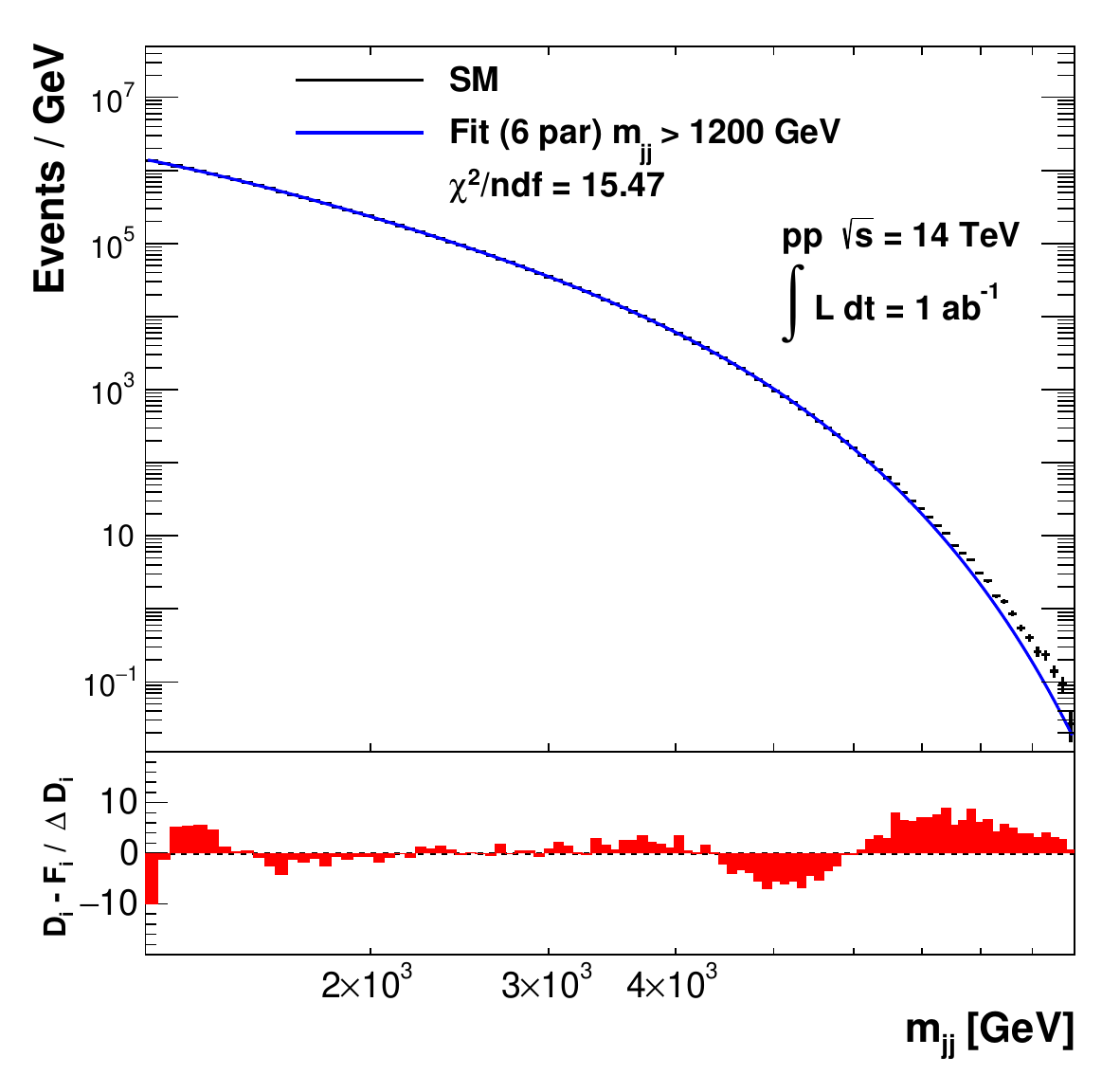}
    }
  \end{center}
  \caption{
    The distributions of the dijet invariant mass of the simulated SM events for the \lumismall and \lumilarge scenarios and functional fits using \Eqn{\ref{eq:bkg}} after adding the additional term $p_6\ln^3 x$ to the exponent of the power law. 
    The bottom panels show the significance of deviations of the fit.
    A ``wave'' pattern for the deviations is observed for the 1~ab$^{-1}$ case 
    indicating failures of the fit to describe the simulated SM events.
  }
  \label{fig:mass_jj_beforecut_p6}
\end{figure*}

\FloatBarrier
\newpage
\section{Acceptance examples}
\label{A2}

\begin{figure*}[htb]

\begin{center}
\subfloat[$W$ leptonic decays]{\includegraphics[width=0.37\textwidth]{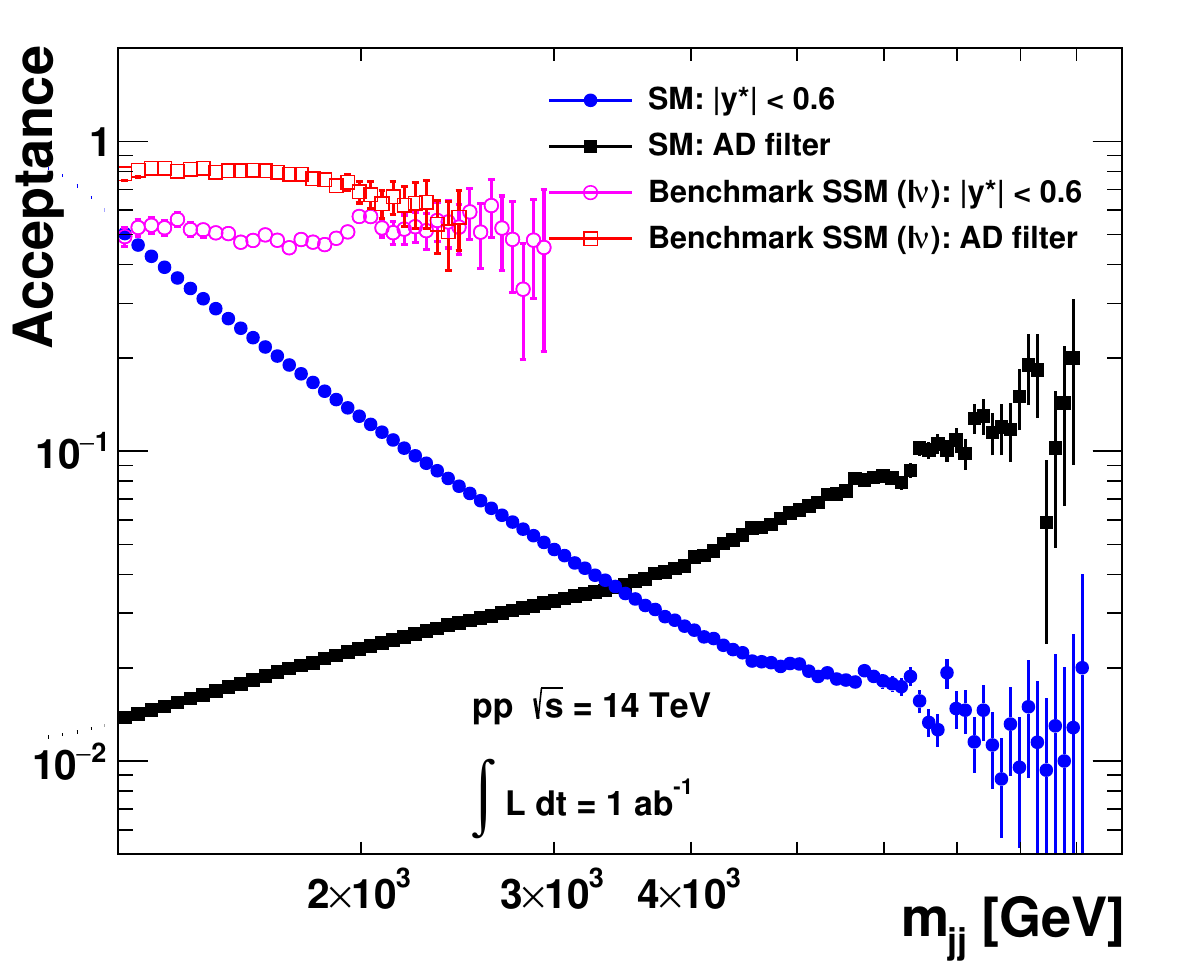}}
\subfloat[$W$ hadronic decays]{
    \includegraphics[width=0.37\textwidth]{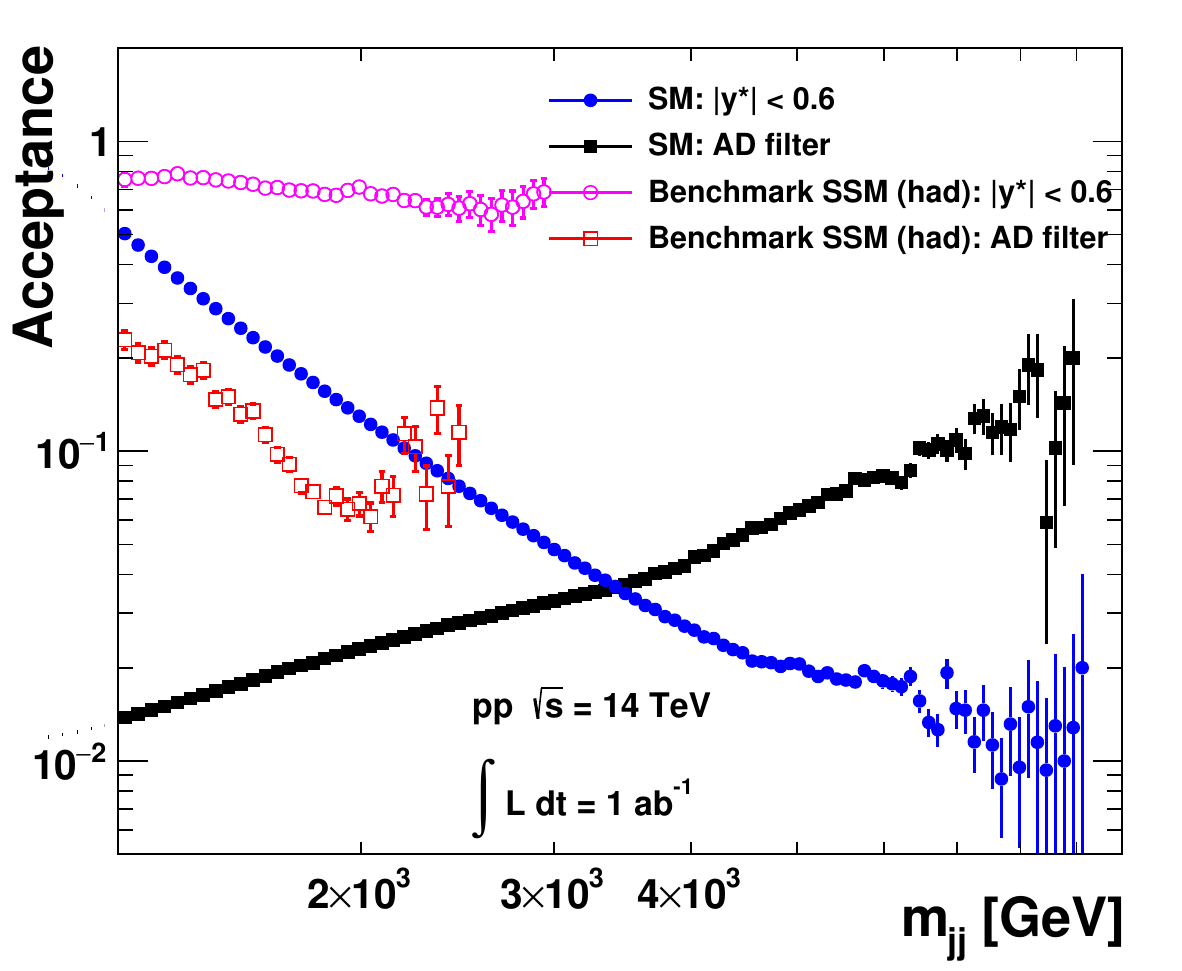}
    }

    \end{center}
    \caption{
        The acceptances of the \ycut selection or the AD filter, evaluated for both the SM events or the benchmark SSM events.
        The SSM model was generated with the mass of $W'$ set to 3 TeV. The acceptances for the SSM are shown for (a) the leptonic $W$ decays, and (b) for the hadronic $W$ decays. 
        Acceptance is determined as the ratio of events retained after applying either the \ycut criterion or the AD filter to the total number of events within each specific mass bin.
    }
    \label{fig:mass_jj_acceptance_3tev_decays}
\end{figure*}

\begin{figure*}[htb]
    \begin{center}{\includegraphics[width=0.5\textwidth]{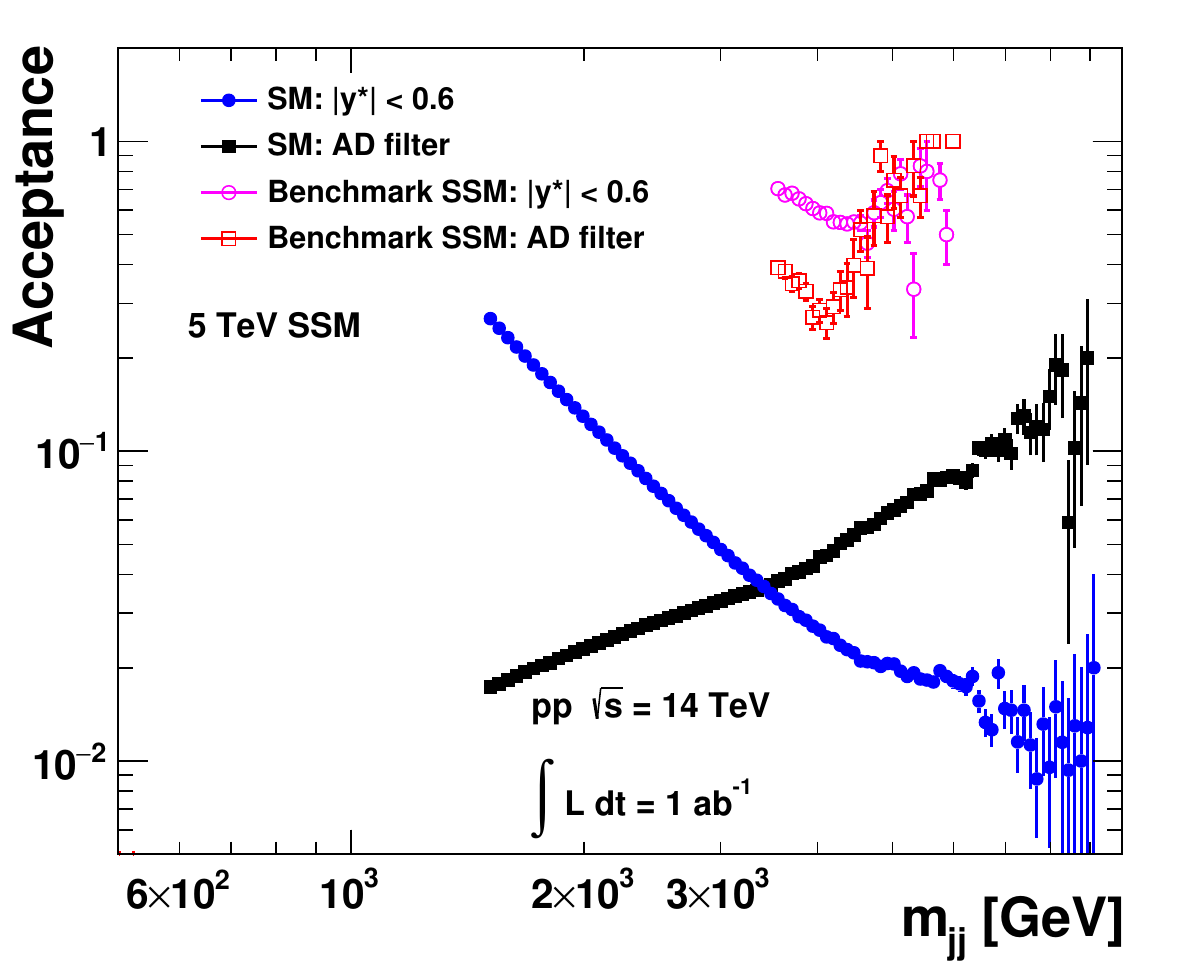}}
    \end{center}
    \caption{
        The acceptance of the \ycut selection or the AD filter, evaluated for both the SM events or the benchmark SSM events.
        The SSM model was generated with the mass of $W'$ set to 5 TeV.
        Acceptance is determined as the ratio of events retained after applying either the \ycut criterion or the AD filter to the total number of events within each specific mass bin.
        Compared to Fig.~\ref{fig:mass_jj_acceptance_3tev_decays}, the $x$-range of this figure is shifted to a lower value in order to fit the legend. 
    }
    \label{fig:mass_jj_acceptance_5tev}
\end{figure*}

\end{document}